\documentclass[10pt,journal,compsoc]{IEEEtran}
\makeatletter
\newif\if@restonecol
\makeatother

\ifCLASSOPTIONcompsoc
  \usepackage[nocompress]{cite}
\else
  \usepackage{cite}
\fi
\usepackage{graphicx}
\usepackage{amssymb}
\usepackage{amsmath}
\usepackage{enumerate}
\usepackage{algorithmic}
\usepackage[linesnumbered,ruled,vlined]{algorithm2e}
\usepackage[usenames]{color}
\usepackage{url}
\usepackage{multirow} 
\usepackage{booktabs}
\usepackage{threeparttable}

\ifCLASSINFOpdf
\else
\fi
\begin{document}
%
\title{When data mining meets optimization: A case study on the quadratic assignment problem}
%
%
%
%

\author{Yangming~Zhou,
        Jin-Kao~Hao$^*$,
        and~B{\'e}atrice~Duval 
\IEEEcompsocitemizethanks{\IEEEcompsocthanksitem Y. Zhou, J.K. Hao (corresponding author) and B. Duval are with the Department of Computer Science, LERIA, Universit\'{e} d'Angers, 2 Boulevard Lavoisier, 49045 Angers, France, J.K. Hao is also affiliated with the Institut Universitaire de France, 1 rue Descartes, 75231 Paris, France (E-mails: zhou.yangming@yahoo.com; jin-hao.hao@univ-angers.fr; beatrice.duval@univ-angers.fr).\protect\\
}
}
\IEEEtitleabstractindextext{%
\begin{abstract}
This paper presents a hybrid approach called frequent pattern based search that combines data mining and optimization. The proposed method uses a data mining procedure to mine frequent patterns from a set of high-quality solutions collected from previous search, and the mined frequent patterns are then employed to build starting solutions that are improved by an optimization procedure. After presenting the general approach and its composing ingredients, we illustrate its application to solve the well-known and challenging quadratic assignment problem. Computational results on the 21 hardest benchmark instances show that the proposed approach competes favorably with  state-of-the-art algorithms both in terms of solution quality and computing time.
\end{abstract}

\begin{IEEEkeywords}
Frequent pattern mining, heuristic design, learning-based optimization, quadratic assignment
\end{IEEEkeywords}}

\maketitle

\IEEEdisplaynontitleabstractindextext

%
\IEEEpeerreviewmaketitle

\IEEEraisesectionheading{\section{Introduction}\label{Sec:Introduction}}

\IEEEPARstart{R}{ecent} years, hybridization of data mining techniques with metaheuristics has received increasing attention from the optimization community. A number of attempts have been made to use data mining techniques to enhance the performance of metaheuristic optimization for solving hard combinatorial problems \cite{Samorani2012,Wauters2013,Asta2015,Wauters2015,Zhou2016,Zhou2017a}. 

Data mining involves discovering useful rules and hidden patterns from data \cite{Witten2005}. With the help of data mining techniques, heuristic search algorithms can hopefully make their search strategies more informed and thus improve their search performances. For instance, candidate solutions visited by a search algorithm can be collected and exploited by supervised or unsupervised learning techniques to extract useful hidden patterns and  rules, which can be used to guide future search decisions of the search algorithm. More generally, data mining techniques could help to build the starting solution or initial population, choose the right operators, set the suitable parameters or make automatic algorithm configuration \cite{Jourdan2006,StutzleLopez2015}. 

In this work, we introduce a hybrid approach called frequent pattern based search (FPBS) that combines data mining techniques and optimization methods for solving combinatorial search problems. Basically, FPBS employs a data mining procedure to mine useful patterns that frequently occur in high-quality (or elite) solutions collected from previous search and then uses the mined patterns to construct new starting solutions that are further improved by an optimization method. To update the set of elite solutions used for pattern mining, FPBS calls for a dedicated procedure to manage newly discovered elite solutions. The key intuition behind the proposed approach is that a frequent pattern extracted from high-quality solutions identifies a particularly promising region in the search space that is worthy of an intensive examination by the optimization procedure. By using multiple mined patterns combined with an effective optimization procedure, FPBS is expected to achieve a suitable balance between search exploration and exploitation that is critical for a high performance of the search algorithm \cite{Hao2012}.

To verify the interest of the proposed FPBS approach, we consider the well-known and highly challenging quadratic assignment problem (QAP) as a case study. Besides its popularity as one of the most studied NP-hard combinatorial optimization problem, QAP is also a relevant representative of many permutation problems. To apply the general FPBS approach to solve QAP, we specify the underlying patterns and frequent patterns, the frequent pattern mining algorithm and the dedicated optimization procedure. We then assess the resulting algorithm on the set of the hardest QAP benchmark instances from the QAPLIB. Our experimental results show that the proposed algorithm is highly competitive compared to state-of-the-art algorithms both in terms of solution quality and computing time. 

The rest of this paper is organized as follows. In the next section, we introduce the concept of frequent pattern mining and provide a  literature review on hybridization of association rule mining with metaheuristic optimization. In Section \ref{Frequent Pattern Based Search}, we present the proposed frequent pattern based search approach. Section \ref{Sec:FPBS Applied to Quadratic Assignment Problem} shows the application of the general FPBS approach to solve the quadratic assignment problem. Section \ref{Sec:Analysis and Discussion} is dedicated to an experimental analysis of key components of the proposed approach. Finally, conclusions and further work are discussed in Section \ref{Sec:Conclusions and Further Work}.

\section{Frequent pattern mining and heuristic search: state of the art}
\label{Sec:Frequent Pattern Mining and Heuristic Search}

In this section, we briefly recall the concept of frequent pattern mining and review related literature on combination of search methods with association rule mining (or frequent pattern mining) techniques for solving combinatorial optimization problems.
%

\subsection{Frequent pattern mining}
\label{SubSec:Frequent Pattern Mining}

Frequent pattern mining was originally introduced for market basket analysis in the form of association rules mining \cite{Agrawal1993a}, with the purpose of identifying associations between different items that customers place in their ``shopping baskets''. Also, the concept of frequent itemsets was first introduced for mining transaction database. Let $\mathcal{D} = \{T_1 ,T_2 ,\ldots,T_N\}$ be a transaction database defined over a set of items $I = \{x_1 ,x_2 ,x_3 ,\ldots,x_M\}$. A frequent itemset typically refers to a set of items that often appear together in a transactional dataset \cite{Han2011}, e.g., milk and bread, which are frequently bought together in grocery stores by many customers. Furthermore, a $l$-itemset (or $l$-pattern) $p \subseteq I$, which consists of $l$ items from $I$, is frequent if $p$ occurs in the transaction database $\mathcal{D}$ no lower than $\theta$ times, where $\theta$ is a user-defined minimum support threshold.

In the original model of frequent pattern mining \cite{Agrawal1993a}, the problem of finding association rules has also been proposed. Association rules are closely related to frequent patterns in the sense that association rules can be considered as a ``second-stage'' output, which are derived from the mined frequent patterns. Given two itemsets $U \subseteq I$ and $V \subseteq I$, the rule $U \Rightarrow V$ is considered to be an association rule at minimum support $\theta$ and minimum confidence $\eta$, when the following two conditions are satisfied simultaneously: the set $U \cup V$ is a frequent pattern for the minimum support $\theta$ and the ratio of the support of $U \cup V$ over $U$ is at least $\eta$.

\subsection{Association rule (or frequent pattern) mining for heuristic search}
\label{SubSec:Association Rule Mining for Heuristic Search}

We now review existing studies on hybridization between data mining and heuristic search, starting by the issue of representation of frequent patterns.

\subsubsection{Representation of frequent patterns}
\label{SubSubSec:Representation of the Frequent Patterns}

The mined knowledge or information can be represented in the form of frequent patterns or association rules. Frequent patterns are patterns that occur frequently in the given data \cite{Han2011}. There are various types of patterns, such as itemsets, subsequences, and substructures. 

To apply frequent pattern mining for solving combinatorial optimization problems, one of the main challenges is to define a suitable pattern for the problem under consideration. In the following, we introduce the definition of frequent patterns (in terms of frequent items) for two categories of representative optimization problems.

\begin{itemize}
	\item Frequent pattern for \textbf{subset selection problems}. Subset selection is basically to determine a subset of specific elements among a set of given elements while optimizing an objective function defined over the selected elements or unselected elements. Subset selection is encountered in many situations and includes, for instance, knapsack problems, clique problems, diversity problems, maximum stable problems and critical node problems. A candidate solution to a subset selection problem is usually represented by a set of selected elements. Since it is natural to treat each element as an item, frequent pattern mining techniques can be directly applied to subset selection problems. In this setting, a transaction corresponds to a solution of the subset selection problem and a pattern can be conveniently defined as a subset of elements that frequently appear in some specific (e.g., high-quality) solutions. Thus, given a database $\mathcal{D}$ (i.e., a set of visited solutions), a frequent pattern $p$ mined from $\mathcal{D}$ with support $\theta$ corresponds to a set of elements that occur at least $\theta$ times in the database.
	\item Frequent pattern for \textbf{permutation problems}. Permutation problems cover another large range of important combinatorial problems. Many classic NP-hard problems, such as the traveling salesman problem, the quadratic assignment problem, graph labeling problems and flow shop scheduling problems, are typical permutation examples. Even if the solutions of this category of problems are   represented as permutations, the practical meaning of a permutation depends on the problem under consideration. To apply frequent pattern mining techniques to such a problem, a transformation between a permutation and an itemset is necessary. For instance, in the context of a vehicle routing problem, a transformation was proposed to map a permutation (a sequence of visited cities) into a set of pairs between any two \textit{consecutive} elements (cities) in the permutation \cite{Guerine2016}. This transformation emphasizes the order between two consecutive elements. In Section \ref{SubSubSec:Mine Frequent Patterns for the QAP}, we introduce a new transformation for QAP whose main idea is to decompose a permutation into a set of element-position pairs, which focuses on not only the order between elements but also the relation between an element and its location. In these settings, a frequent pattern can be considered as a set of pairs that frequently appear in some specific solutions. 
\end{itemize}

\begin{table*}[!htbp]
\scriptsize\centering
\caption{Main work on hybridizing association rules mining techniques with metaheuristics for solving combinatorial optimization problems.}
\label{Tab:Literature Review on Hybrid Algorithms}
\begin{threeparttable}
\begin{tabular}{lllc}
\toprule[0.75pt]
Algorithm 		    & metaheuristic\tnote{$\star$}  & optimization problem & year\\
\midrule[0.5pt]
DM-GRASP \cite{Ribeiro2004,Ribeiro2006} & GRASP & set packing problem & 2004,2006\\
DM-GRASP \cite{Santos2005} & GRASP & maximum diversity problem & 2005\\
DM-GRASP \cite{Santos2006a}& GRASP & server replication for reliable multicast problem & 2006\\
GADMLS \cite{Santos2006b}   & GA+LS & single-vehicle routing problem & 2006\\
DM-GRASP \cite{Plastino2011}& GRASP & p-median problem & 2011\\
HDMNS \cite{Reddy2012}       & NS    & p-median problem & 2012\\
DM-GRASP-PR \cite{Barbalho2013} & GRASP+PR & 2-path network design problem &2013\\
MDM-GRASP \cite{Plastino2014} & GRASP & server replication for reliable multicast problem & 2014\\
DM-HH \cite{Martins2014} & PR+LS& p-median problem & 2014\\
VALS \cite{Umetani2015} & LS & set partitioning problem & 2015\\
GAAR \cite{Raschip2015a} & GA & constrain satisfaction problem & 2015\\
GAAR \cite{Raschip2015b} & GA & weighted constrain satisfaction problem & 2015\\
MDM-GRASP/VND \cite{Guerine2016} & GRASP+VND & one-commodity pickup-and-delivery traveling salesman problem & 2016\\
DM-ILS \cite{Martins2016} & ILS & set covering with pairs problem &2016\\
\bottomrule[0.75pt]
\end{tabular}
\begin{tablenotes}
     \item[$\star$] Greedy randomized adaptive search procedure (GRASP), genetic algorithm (GA), local search (LS), neighborhood search (NS), path-relinking (PR), variable neighborhood descent (VND), and iterated local search (ILS).
\end{tablenotes}
\end{threeparttable}
\end{table*}

\subsubsection{Mining and heuristics}

In this section, we present a brief literature review on hybridizing association rule mining (or frequent pattern mining) techniques with metaheuristics. The reviewed studies are summarized in Table \ref{Tab:Literature Review on Hybrid Algorithms}.

GRASP (Greedy randomized adaptive search procedure) was the first metaheuristic to be hybridized with data mining techniques (i.e., association rule mining) \cite{Ribeiro2004,Ribeiro2006}. Denoted as DM-GRASP, this approach was originally designed to solve the set packing problem. DM-GRASP organizes its search process into two sequential phases, and incorporates an association rule mining procedure at the second phase. Specifically, high-quality solutions found in the first phase (GRASP) were stored in an elite set, and then a data mining procedure is invoked to extract some patterns from the elite set. At the second phase, a new solution is constructed based on a mined pattern instead of using the greedy randomized construction procedure of GRASP. This approach has been applied to solve several problems including the maximum diversity problem \cite{Santos2005}, the server replication for reliable multicast problem \cite{Santos2006a}, and the p-median problem \cite{Plastino2011}. A survey on some significant applications of DM-GRASP can be found in \cite{Santos2008}.  An interesting extension of DM-GRASP is to execute the data mining procedure multiple times instead of only once \cite{Plastino2014,Guerine2016,Martins2016}. Compared to DM-GRASP where the data mining call occurs once at the midway of the whole search process, the multi-mining version performs the mining task when the elite set stagnates. The same idea has been explored recently by hybridizing data mining and GRASP enhanced with path-relinking (PR) \cite{Barbalho2013} or variable neighborhood descent (VND) \cite{Guerine2016}. 

In addition to GRASP, data mining has also been hybridized with other metaheuristics like evolutionary algorithms. To improve the performance of an evolutionary algorithm applied to an oil collecting vehicle routing problem, a hybrid algorithm (GADMLS) combining genetic algorithm, local search and data mining was proposed in \cite{Santos2006b}. Another hybrid approach (GAAR) that uses a data mining module to guide an evolutionary algorithm was presented in \cite{Raschip2015a,Raschip2015b} to solve the constraint satisfaction problem (CSP). Besides the standard components of a genetic algorithm, a data mining module is added to find association rules (between variables and values) from an archive of best individuals found in the previous generations. 

Apart from GRASP and evolutionary algorithms, it has been shown that other heuristics can also benefit from the incorporation of a data mining procedure. For example, a data mining approach was applied to extract variable associations from previously solved instances for identifying promising pairs of flipping variables in a large neighborhood search method for the set partitioning problem, thus reducing the search space of local search algorithms for the set partitioning problem \cite{Umetani2015}. Another example is the hybridization of  neighborhood search with data mining techniques for solving the p-median problem \cite{Reddy2012}. A data mining procedure was also integrated into a multistart hybrid heuristic for the p-median problem \cite{Martins2014}, which combines elements of different traditional metaheuristics (e.g., local search) and uses path-relinking as a memory-based intensification mechanism. Finally, the widely-used iterated local search method was recently hybridized with data mining for solving the set covering with pairs problem \cite{Martins2016}. 

\section{Frequent pattern based search}
\label{Frequent Pattern Based Search}

In this section, we propose frequent pattern based search (FPBS) for solving combinatorial search problems. This general-purpose search approach tightly integrates frequent pattern mining with an optimization procedure. Below, we first show the general scheme of the proposed FPBS approach, and then present its composing ingredients.

\subsection{General scheme}
\label{SubSec:General Scheme}

The FPBS approach uses relevant frequent patterns extracted from high-quality solutions to build promising starting solutions which are further improved by an optimization procedure. The improved solutions are in turn used to help mine additional patterns. By iterating the mining procedure and the optimization procedure, FPBS is expected to examine the search space effectively and efficiently. As a case study, we illustrate in Section \ref{Sec:FPBS Applied to Quadratic Assignment Problem} its application to solve the well-known quadratic assignment problem. 

From a perspective of system architecture, FPBS maintains an archive of high-quality solutions (called elite set) for the purpose of pattern mining and includes five critical operating components: an initialization procedure (Section \ref{SubSec:Elite Set Initialization}), an optimization search procedure (Section \ref{SubSec:Local Optimization Search}), a data mining procedure (Section \ref{SubSec:Frequent Pattern Mining Procedure}), a frequent pattern based solution construction procedure (Section \ref{SubSec:Construction Based on Mined Pattern}) and an elite solution management procedure (Section \ref{SubSec:Elite Set Management Strategy}). 

\begin{algorithm}[ht]
\begin{small}
 \caption{General framework of Frequent Pattern Based Search}
 \label{Alg:FPBS Framework}
\KwIn{Problem instance $I$, elite set size $k$ and number of patterns to be mined $m$}
\KwOut{The best solution $S^*$ found}
\Begin{
 	// $I$ is supposed to be a minimization problem
 	\\
 	$ES \leftarrow EliteSetInitialize()$; /* $ES$ is the set of elite solution */\\
 	$S^* \leftarrow \arg \min \{f(S_i): i = 1,2,\ldots,k\}$; /* $S^*$ is the best solution found so far */\\
 	$\mathcal{P} \leftarrow FrequentPatternMine(ES,m)$;\\
	\While{a stopping condition is not reached}{
		$p \leftarrow PatternSelection(\mathcal{P})$;\\ 
		// construct a new solution based mined frequent pattern
		$S \leftarrow PatternBasedConstruct(p)$;\\
		// improve the constructed solution
		\\
		$S' \leftarrow Optimize(S)$;\\
		// update the best solution found so far
		\\
		\If{$f(S') < f(S^*)$}{
			$S^* \leftarrow S'$;
			}
		// update the elite set
		\\
		$ES \leftarrow EliteSetUpdate(ES,S')$;\\
		// restart the mining procedure when the elite set stagnates\\
		\If{elite set stagnates}{
			$\mathcal{P} \leftarrow FrequentPatternMine(ES,m)$;
			}
		}
}
\end{small}
\end{algorithm}

The general framework of the proposed FPBS approach is presented in Algorithm \ref{Alg:FPBS Framework}. FPBS starts from a set of high-quality solutions that are obtained by the initialization procedure (line 3). From these high-quality solutions, a data mining procedure is invoked to mine a number of frequent patterns (line 5). A new solution is then constructed based on a mined pattern and further improved by the optimization procedure (lines 7-10). The improved solution is finally inserted to the elite set according to the elite solution management policy (line 15). The process is repeated until a stopping condition (e.g., a time limit) is satisfied. In addition to its invocation just after the initialization procedure, the data mining procedure is also called each time the elite set is judged to be stagnating (lines 17-18), i.e., it has not been updated during a predefined number of iterations.

\subsection{Elite set initialization}
\label{SubSec:Elite Set Initialization}

FPBS starts its search with an elite set ($ES$) composed of $k$ distinct high-quality solutions. To build such an elite set, we first generate, by any means (e.g., with a random or greedy construction method), an initial solution that is improved by an optimization procedure (see Section \ref{SubSec:Local Optimization Search}). The improved solution is then inserted into the elite set according to the elite set management strategy (see Section \ref{SubSec:Elite Set Management Strategy}). We repeat this process until an elite set of $k$ different solutions is built. Note that similar ideas have been successfully applied to build a high-quality initial population for memetic algorithms \cite{Lv2010,Benlic2015,Zhou2017a}.

\subsection{Frequent pattern mining procedure}
\label{SubSec:Frequent Pattern Mining Procedure}

In our approach, the frequent pattern mining procedure is used to discover some  patterns that frequently occurs in high-quality solutions stored in the elite set. Beside the itemsets (see Section \ref{SubSubSec:Representation of the Frequent Patterns}), mined patterns can also be subsequences or substructures \cite{Aggarwal2014}. A subsequence is an order of items, e.g., buying a mobile phone first, then a power bank, and finally a memory card. If a subsequence occurs frequently in a transaction database, then it is a frequent sequential pattern. A substructure can refer to different structural forms, such as subgraphs, subtrees, or sublattices, which may be combined with itemsets or subsequences. If a substructure occurs frequently in a graph database, it is called a frequent structural pattern.

\begin{table*}[!ht]
\scriptsize\centering
\caption{A simple summary of frequent pattern mining algorithms \cite{Aggarwal2014}.}
\label{Tab:Classification On Mining Tasks}
\begin{tabular}{llll}
\toprule[0.75pt]
\multicolumn{2}{c}{} & \multicolumn{2}{c}{algorithms}\\
\cmidrule[0.5pt]{3-4}
tasks & patterns & apriori-based & pattern-growth \\
\midrule[0.5pt]
frequent itemset mining & itemsets & e.g., Apriori & e.g., FP-growth, FPmax\\
sequential pattern mining & subsequences & e.g., GSP, SPADE & e.g., PrefixSpan\\
structural pattern mining & substructures & e.g., AGM, FSG & e.g., gSpan, FFSM\\
\bottomrule[0.75pt]
\end{tabular}
\end{table*}

To handle a wide diversity of data types, numerous mining tasks and algorithms have been proposed in the literature \cite{Grahne2005,Han2007,Aggarwal2014}. A simple summary of frequent pattern mining algorithms is provided in Table \ref{Tab:Classification On Mining Tasks}. For a specific application, the patterns of high-quality solutions collected in the elite set can be expressed as itemsets, subsequences, or substructures. Once the form of patterns is determined, a suitable mining algorithm can be selected accordingly. For our case study on QAP presented in Section \ref{Sec:FPBS Applied to Quadratic Assignment Problem}, a pattern corresponds to a set of element-location pairs. Thus frequent patterns can be conveniently represented as frequent itemsets (i.e., a set of identical element-location assignments). Consequently, we adopt the FPmax* algorithm (see Section \ref{SubSubSec:Mine Frequent Patterns for the QAP} for more details) to mine only the maximal frequent items. Detailed reviews on frequent pattern mining algorithms can be found in \cite{Han2007,Aggarwal2014}.

\subsection{Optimization procedure}
\label{SubSec:Local Optimization Search}

For the purpose of solution improvement (to built the initial elite set and to improve each new solution built from a mined pattern), any optimization procedure dedicated to the given problem can be applied in principle. On the other hand, since the optimization component ensures the key role of search intensification, it is desirable to call for a powerful search algorithm. Basically, the optimization procedure can be considered to be a black box optimizer that is called to improve the input solution. In practice, the optimization procedure can be based on local search, population-based search or even hybrid memetic search. In any case, the search procedure must be carefully designed with respect to the problem under consideration and should ideally be effective both in terms of search capacity and time efficiency. As we show in Section \ref{Sec:FPBS Applied to Quadratic Assignment Problem}, for the quadratic assignment problem considered in this work, we will adopt the powerful breakout local search procedure \cite{Benlic2013a} as our optimization procedure. 

\subsection{Solution construction based on mined pattern}
\label{SubSec:Construction Based on Mined Pattern}

Once a set of frequent patterns $\mathcal{P}$ is extracted from the elite set, new solutions are constructed based on these mined  patterns. For this purpose, we first select a mined frequent pattern by the tournament selection strategy as follows. Let $\lambda$ be the size of the tournament pool. We randomly choose $\lambda~(1 \leqslant \lambda \leqslant |\mathcal{P}|)$ individuals with replacement from the mined pattern set $\mathcal{P}$, and then pick the best one (i.e., with the largest size), where $\lambda$ is a parameter. The computational complexity of this selection strategy is $O(|\mathcal{P}|)$. The advantage of the tournament selection strategy is that the selection pressure can be easily adjusted by changing the size of the tournament pool $\lambda$. The larger the tournament pool is, the smaller the chance for shorter patterns to be selected.

Since frequent patterns usually correspond to a set of common elements shared by the high-quality solutions examined by the mining procedure, each mined pattern directly defines a partial solution. To obtain a whole solution, we can apply a greedy or random procedure to complete the partial solution. The way to build such a solution shares similarity with the general backbone-based crossover procedure \cite{Benlic2011,Zhou2017a}. However, compared to the notion of backbones that are typically shared by two parent solutions, our frequent patterns are naturally shared by two or more high-quality solutions. In this sense, backbones can be considered as a special case of more general frequent patterns.

Finally, it is possible to construct a new solution for each mined pattern instead of using a long pattern selected by the tournament selection strategy, as explained above. Also, an elite solution can be selected to guide the construction of a new solution. In particular, the way to use frequent patterns should be determined according to the studied problem. 

\subsection{Elite set management}
\label{SubSec:Elite Set Management Strategy}

As explained above, each new solution constructed using a mined frequent pattern is improved by the optimization procedure. Then, we decide whether the improved solution should be inserted into the elite set $ES$. 

There are a number of updating strategies in the literature \cite{Sorensen2006} that can be applied within the FPBS approach. For example, the classic quality-based replacement strategy simply inserts the solution into $ES$ to replace the worst solution if it is better than the worst solution in $ES$. In addition, more elaborated updating strategies consider other criteria than the quality of solutions. For example, the quality-and-distance updating strategy not only considers the quality of the solution, but also its distance to other solutions in the population \cite{Lv2010,Porumbeletal2010,Zhou2017a}. A suitable elite set management strategy can be determined according to the practical problem. 

\section{FPBS applied to the quadratic assignment problem}
\label{Sec:FPBS Applied to Quadratic Assignment Problem}

In this section, we present a case study of applying the general FPBS approach to the well-known quadratic assignment problem (QAP) and show its competitiveness compared to state-of-the-art QAP algorithms. 

\subsection{Quadratic assignment problem}
\label{SubSec:Quadratic Assignment Problem}

The quadratic assignment problem is a well-known NP-hard combinatorial optimization problem. It was originally introduced by Koopmans and Beckman \cite{Koopmans} in 1957 to model the locations of indivisible economic activities such as capital equipment. QAP aims to determine a minimal cost assignment of $n$ facilities to $n$ locations, given a flow $a_{ij}$ from facility $i$ to facility $j$ for all $i,j \in \{1,\ldots,n\}$ and a distance $b_{uv}$ between locations $u$ and $v$ for all $u,v \in \{1,\ldots,n\}$. Let $\Omega$ denotes the set of all possible permutations $\pi:\{1,\ldots,n\} \rightarrow \{1,\ldots,n\}$, then QAP can mathematically be formulated as follows.

\begin{equation}\label{Equ:QAP Objective Function}
	\min_{\pi \in \Omega} f(\pi) = \sum^n_{i=1} \sum^n_{j=1} a_{ij}b_{\pi(i) \pi(j)}
\end{equation}

where $a$ and $b$ are the flow and distance matrices respectively, and $\pi \in \Omega$ is a solution and $\pi(i)$ represents the location chosen for facility $i$. The optimization objective is to find a permutation $\pi^*$ in $\Omega$ such that the sum of the products of the flow and distance matrices, i.e., $f(\pi^*) \leqslant f(\pi), \forall \pi \in \Omega$, is minimized.

In addition to the facility location application, QAP can be used to formulate many other real-world problems \cite{Drezner2005,Loiola2007,Duman2007} such as electrical circuit wiring/routing, transportation engineering, parallel and distributed computing, image processing and analysis of chemical reactions for organic compounds. Moreover, a number of classic NP-hard problems, such as the traveling salesman problem, the maximum clique problem, the bin packing problem and the graph partitioning problem, can also be recast as QAPs \cite{Loiola2007}.

Due to its practical and theoretical significance, QAP has attracted much research effort since its first formulation \cite{James2009,Benlic2013a,Benlic2015,Tosun2015,Acan2015}. In fact, QAP is among the most studied and the most competitive combinatorial optimization problems. Since exact algorithms are unpractical for instances of size larger than 36 \cite{Anstreicher2002}, a large number of heuristic methods have been proposed for QAP, which can provide near-optimal solutions in reasonable computation times. Detailed reviews of heuristic and metaheuristic algorithms developed till 2007 for QAP are available in \cite{Drezner2005,Loiola2007}. An updated review of more recent studies on QAP can be found in \cite{Benlic2013a,Benlic2015}.

\subsection{FPBS for QAP}
\label{SubSec:FPBS for QAP}

Algorithm \ref{Alg:FPBS-QAP} shows the FPBS algorithm for QAP (denoted as FPBS-QAP), which is an instantiation of the general scheme of Algorithm \ref{Alg:FPBS Framework}. Since FPBS-QAP inherits the main components of FPBS, hereafter we only present the specific features related to QAP: solution representation and evaluation, optimization procedure, frequent pattern mining for QAP, solution construction using QAP patterns, and elite set update strategy.

\begin{algorithm}[ht]
\begin{small}
 \caption{The FPBS algorithm for QAP.}
 \label{Alg:FPBS-QAP}
\KwIn{Instance G, elite set size $k$, the number of mined patterns $m$, time limit $t_{max}$ and the maximum number of iterations without updating $max\_no\_update$}
\KwOut{The best solution $\pi^*$ found so far}
\Begin{
	$ES \leftarrow EliteSetInitialize()$;\\
	$\pi^* \leftarrow \arg \min \{f(\pi_i): i = 1,2,\ldots,k\}$;\\
	$\mathcal{P} \leftarrow FrequentPatternMine(ES,m)$;\\
	$no\_update \leftarrow 0$;\\
	$t \leftarrow 0$;\\
	\While{$t < t_{max}$}{
		$p_i \leftarrow PatternSelection(\mathcal{P})$;\\
		// build a new solution based selected pattern\\
		$\pi \leftarrow PatternBasedConstruct(p_i)$;\\
		// improve the constructed solution
		\\
		$\pi' \leftarrow BreakoutLocalSearch(\pi)$;\\
		// update the best solution found so far
		\\
		\If{$f(\pi') < f(\pi^*)$}{
			$\pi^* \leftarrow \pi'$;
			}
		// update the elite set
		\\
		\If{$EliteSetUpdate(ES,\pi') = True$}{
			$no\_update \leftarrow 0$;
			}
		\Else{
			$no\_update \leftarrow no\_update + 1$;
			}
		// restart the mining procedure when the elite set is steady
		\\
		\If{$no\_update > max\_no\_update$}{
			$\mathcal{P} \leftarrow FrequentPatternMine(ES,m)$;\\
			$no\_update \leftarrow 0$;
			}
	}
}
\end{small}
\end{algorithm}

\subsubsection{Solution representation, neighborhood and evaluation}
\label{SubSubSec:Solution Representation and Evaluation}

Given a QAP instance with $n$ facilities (or locations), a candidate solution is naturally represented by a permutation $\pi$ of $\{1,2,\ldots,n\}$, such that $\pi(i)$ is the location assigned to facility $i$. The search space $\Omega$ is thus composed of all possible $n!$ permutations. For any solution $\pi \in \Omega$, its quality is given by Eq. (\ref{Equ:QAP Objective Function}).

To examine the search space, we adopt an iterated local search algorithm called BLS (see Section \ref{SubSubSec:Breakout Local Search}). For this purpose, we introduce the neighborhood used by BLS. Give a solution, i.e., a permutation $\pi$, its neighborhood $N(\pi)$ is defined as the set of all possible permutations that can be obtained by exchanging the values of any two different positions $\pi(u)$ and $\pi(v)$ in $\pi$, i.e., $N(\pi) = \{\pi'~|~\pi'(u) = \pi(v), \pi'(v) = \pi(u), u \neq v~and~\pi'(i) = \pi(i), \forall i \neq u,v\}$, which has a size of $n(n-1)/2$.

As indicated in \cite{Taillard1991}, given a permutation $\pi$ and its objective value $f(\pi)$, the objective value of any neighnoring permutation $\pi'$ can be effectively calculated according to an incremental evaluation technique. 

\subsubsection{Breakout local search}
\label{SubSubSec:Breakout Local Search}

To ensure an effective examination of the search space, we adopt the breakout local search (BLS) algorithm \cite{Benlic2013a}, which is a state-of-the-art algorithms for QAP currently available in the literature. 

BLS follows the iterated local search scheme and iteratively alternates between a descent search phase (to find local optima) and a dedicated perturbation phase (to discover new promising regions). BLS starts from an initial random permutation, and then improves the initial solution to a local optimum by the best improvement descent search with the above exchange-based neighborhood. Upon the discovery of a local optimum, BLS triggers a perturbation mechanism. The perturbation mechanism adaptively selects a tabu-based perturbation (called directed perturbation) or a random perturbation (called undirected perturbation). BLS also determines the number of perturbation steps (called perturbation strength) in an adaptive way \cite{Benlic2013a}. 

The tabu-based perturbation and random perturbation provide two complementary means for search diversification. The former applies a selection rule that favors neighboring solutions that minimize the objective degradation, under the constraint that the neighboring solutions have not been visited during the last $\gamma$ iterations (where $\gamma$ is the pre-defined tabu tenure), while the latter performs moves selected uniformly at random. To keep a suitable balance between an intensified search and a diversified search, BLS alternates probabilistically between these two perturbations. The probability to select a particular perturbation is determined dynamically according to the current number of times to visit local optima without improvement on the best solution found. The probability of applying the tabu-based perturbation over the random perturbation is empirically limited to be at least as $Q$. The perturbation strength $L$ is determined based on a simple reactive strategy. One increases $L$ by one if the search returns to the immediate previous local optimum, and otherwise resets $L$ to a given initial value $L_0$. Once the type and the strength $L$ of the perturbation are determined, the selected perturbation with strength $L$ is applied to the current solution. The resulting solution is then used as the starting solution of the next round of the descent search procedure (see \cite{Benlic2013a} for more details).

\subsubsection{Mining frequent patterns for QAP}
\label{SubSubSec:Mine Frequent Patterns for the QAP}

The quadratic assignment problem is a typical permutation problem whose solutions are naturally represented by permutations. For QAP, we define a frequent pattern to be a set of identical location-facility assignments shared by high-quality solutions, and represent a frequent pattern by an itemset. To apply a frequent itemset mining algorithm, we need to transform a permutation into a set of items. A transformation is recently proposed in \cite{Guerine2016}. The transformation works as follows. For each pair of elements ($\pi(i)$ and $\pi(j)$) of a given permutation $\pi$, an arc $(\pi(i),\pi(j))$ is generated, which maps a permutation $\pi$ to a set $S'$ of $|\pi|-1$ arcs. For example, considering a permutation $\pi = \{5,4,7,2,1,6,3\}$ represented by a sequence of these elements, $\pi$ can be mapped as $S' = \{(5,4),(4,7),(7,2),(2,1),(1,6),(6,3)\}$. By this transformation, a permutation is divided into a set of arcs, which conserves  the order of elements. However, this transformation loses the information between the elements and their locations. In practice, we can not identify the true location of an element when only a part of pairs are available.

To overcome this difficulty, we propose a new transformation. Our transformation decomposes a permutation into a set of ordered pairs, each pair being formed by an element (facility) $i$ and its position (location) $\pi(i)$. Specifically, let $\pi$ be a candidate solution of QAP, for each element (facility) $i$, a corresponding element-position pair $(i,\pi(i))$ is generated, which transforms permutation $\pi$ into a set of $n$ element-position pairs. For example, given $\pi = \{5,4,7,2,1,6,3\}$, $\{(1,5),(2,4),(3,7),(4,2),(5,1),(6,6),(7,3)\}$ is the corresponding set of element-position pairs. 

Once a permutation is transformed into a set of pairs, we will treat each pair $(x,y)$ as an item $z = x*n+y$, where $n$ is the length of the permutation. Now, the task of mining frequent patterns from multiple permutations can be conveniently transformed into the task of mining frequent itemsets. The main drawback of mining all frequent itemsets is that if there is a large frequent itemset, then almost all subsets of the itemset might be examined. However, it is usually sufficient to find only the maximal frequent itemsets (a maximal frequent itemset is that it has no superset that is frequent). Thus mining frequent itemsets can be reduced to mine only the maximal frequent itemsets. For this purpose, we adopt the popular \textbf{FPmax*}\footnote{The source code of the FPmax* algorithm is publicly available at \url{http://fimi.ua.ac.be/src/}} algorithm \cite{Grahne2003}.

\begin{algorithm}[!ht]
\begin{small}
\caption{Pseudo code of the \textbf{FPmax*} Algorithm \cite{Grahne2003}}
\label{Alg: FPMAX* Algorithm}
\KwIn{$T$: a FP-tree\\
~~~~~~~~~~$M$: the MFI-tree for $T.base$.}
\KwOut{Updated M}
\Begin{
\If{$T$ only contains a single path $P$}{
	insert $P$ into $M$;
	}
\Else{
	\For{each $i$ in $T$.header}{
		set $Y = T.base \cup \{i\}$;\\
		\If{T.array is not $Null$}{
			$Tail$=$\{$frequent items for $i$ in $T.array\}$;
		}
		\Else
		{
			$Tail$=$\{$frequent items in $i$'s conditional pattern base$\}$;
		}
		sort $Tail$ in decreasing order of the items' counts;\\
		\If{subset\_checking$(Y \cup Tail, M) = False$}{
			construct Y's conditional FP-tree $T_Y$ and its array $A_Y$;\\
			initialize Y's conditional MFI-tree $M_Y$;\\
			call \textbf{FPmax*}$(T_Y , M_Y)$;\\
			merge $M_Y$ with $M$;
			}
		}
	}
}
\end{small}
\end{algorithm}

Algorithm \ref{Alg: FPMAX* Algorithm} shows the pseudo code of the \textbf{FPmax*} algorithm. A description of the basic concepts used in FPmax* such as the frequent pattern tree (FP-tree), the FP-growth method, and the maximal frequent itemset tree (MFI-tree) can be found in \cite{Grahne2003}. Here we only provide a brief presentation of its general procedure. In the initial call, a FP-tree $T$ is constructed from the first scan of the database, together with an initial empty MFI-tree. During the recursion, if there is only one single path in the FP-tree $T$, this single path together with $T.base$ is a MFI of the dataset. The MFI is then inserted into $M$ (line 3). Otherwise, for each item $i$ in the header table, we set $Y = T.base \cup \{i\}$ and prepare for the recursive call FPmax*$(T_Y ,M_Y)$. The items in the header table are processed in increasing order of frequency, so that maximal frequent itemsets will be found before any of their frequent subsets (line 11). Lines 7-10 use the array technique, and line 12 invokes the $subset\_checking()$ function to check if $Y$ together with all frequent items in $Y$'s conditional pattern base is a subset of any existing MFI in $M$, thus we perform superset pruning. If function $subset\_checking()$ returns $False$, FPmax* will be called recursively, with $(T_Y ,M_Y)$ (lines 13-16). For a detailed description of the FPmax* algorithm, please refer to \cite{Grahne2003}.

\subsubsection{Solution construction based on mined pattern}
\label{SubSubSec:Solution Construction Based on Mined Pattern}

Algorithm \ref{Alg:Solution Construction Based on Mined Pattern} describes the main steps to construct new solutions based on the mined frequent patterns. Initially, a pattern is first selected from the set of mined frequent patterns $\mathcal{P}$ according to the tournament selection strategy (line 2). Then, we re-map this chosen pattern into a partial solution $\pi$ (line 3). If the length of the partial solution $|\pi|$ is less than a given threshold (i.e., $\beta*n$), we use an elite solution to guide the construction (lines 4-6). Specifically, we first select an elite solution $\pi^0$ (called guiding solution) from the elite set (line 5), and then we complete $\pi$ based on the guiding solution $\pi^0$ (i.e., directly copy the elements of all unassigned positions of $\pi^0$ to $\pi$ if the elements have not been assigned in $\pi$). Finally, if $\pi$ is still an incomplete solution, we randomly assign the remaining elements to the unassigned positions until a full solution is obtained (line 7).

\begin{algorithm}[!h]
\begin{small}
\caption{Solution construction based on mined frequent patterns}
\label{Alg:Solution Construction Based on Mined Pattern}
\KwIn{Instance G, a set of $m$ mined patterns $\mathcal{P}$ and an elite set $ES$ of size $k$}
\KwOut{A new solution $\pi$}
\Begin{
$p \leftarrow PatternSelection(\mathcal{P})$; $/*$ select a mined pattern $*/$\\
$\pi \leftarrow re$-$map(p)$; $/*$ generate a partial solution based on selected pattern $*/$\\
\If{$|\pi| < \beta*n$}{
	$\pi^0 \leftarrow GuidedSolutionSelect(ES)$; $/*$ select a guiding solution $*/$\\
	$\pi \leftarrow GuidedComplete(\pi,\pi^0)$; $/*$ complete based on guiding solution $*/$
}
$\pi \leftarrow RandomComplete(\pi)$; $/*$ complete at random $*/$
}
\end{small}
\end{algorithm}

\subsubsection{Elite set update strategy}
\label{SubSubSec:Elite Set Update Strategy}

Once a new improved solution $\pi'$ is obtained by the BLS algorithm, we decide whether $\pi'$ should be inserted into the elite set $ES$. In our case, we adopt the following quality-based strategy that $\pi'$ is inserted into $ES$ if two conditions are satisfied simultaneously: (i) $\pi'$ is different from any solution in $ES$ and (ii) $\pi'$ is not worse than any solution in $ES$, i.e., $f(\pi') \leqslant f(\pi^w)$, where $\pi^w \leftarrow \arg \max_{\pi \in ES}\{f(\pi)\}$ is the worst solution in the $ES$. 

\subsection{Computational studies of FPBS for QAP}
\label{SubSec:Computational Studies on QAP}

Our computational studies aim to evaluate the efficiency of the FPBS-QAP algorithm. For this purpose, we first perform a detailed performance comparisons between FPBS-QAP and two state-of-the-art algorithms, i.e., BLS \cite{Benlic2013a} and BMA \cite{Benlic2015}, whose source codes are available to us. Furthermore, we compare FPBS-QAP with four additional algorithms that are published very recently since 2015. 

\subsubsection{Benchmark instances}
\label{SubSubSec:Benchmark Instances}

The experimental evaluations of QAP algorithms are usually performed on 135 popular benchmark instances from QAPLIB\footnote{\url{https://www.opt.math.tugraz.at/qaplib/}}, with $n$ ranging from 12 to 150. These instances can be classified into four categories:
\begin{itemize}
	\item Type I. 114 \textbf{real-life instances} are obtained from practical QAP applications;
	\item Type II. 5 \textbf{unstructured, randomly generated instances} whose distance and flow matrices are randomly generated based on a uniform distribution;
	\item Type III. 5 \textbf{real-like-life instances} are generated instances that are similar to the real-life QAP instances;
	\item Type IV. 11 \textbf{instances with grid-based distances} in which the distances are the Manhattan distance between points on a grid.
\end{itemize}

Like \cite{Benlic2013a,Benlic2015}, we do not consider the 114 instances from Type I because they are very easy for modern QAP algorithms (i.e., their optimal solutions can be found easily with a short time, often less than one second). Our experiments focus on the remaining 21 hard instances from Type II, Type III and Type IV. Notice that for these 21 challenging instances, no single algorithm can attain the best-known results for all the instances. Indeed, even the best performing algorithm misses at least two best-known results.

\subsubsection{Experimental settings}
\label{SubSubSec:Experimental Settings}

The proposed FPBS-QAP algorithm\footnote{The source code of our FPBS-QAP algorithm will be made available at \url{http://www.info.univ-angers.fr/~hao/fpbs.html}} was implemented in the C++ programming language and complied with gcc 4.1.2 and flag `-O3'. All the experiments were carried out on a computer equipped with an Intel E5-2670 processor with 2.5 GHz and 2 GB RAM operating under the Linux system. Without using any compiler flag, running the well-known DIMACS machine benchmark procedure dfmax.c\footnote{dfmax: \url{ftp://dimacs.rutgers.edu/pub/dsj/clique}} on our machine requires 0.19, 1.17 and 4.54 seconds to solve the benchmark graphs r300.5, r400.5 and r500.5 respectively. Our computational results were obtained by running the FPBS-QAP algorithm with the parameter settings provided in Table \ref{Tab:Parameter Settings}. To identify an appropriate value for a given parameter, we compare the performance of the algorithm with different parameter values, while fixing other parameter values. An example to select a appropriate $m$ (frequent pattern set size)  value is provided in Section \ref{SubSec:Impact of the Number of the Largest Patterns}. 

\begin{table}[!ht]
\scriptsize\centering
\caption{Parameter settings of FPBS-QAP algorithm.}
\label{Tab:Parameter Settings}
\begin{threeparttable}
\begin{tabular}{lllc}
\toprule[0.75pt]
Parameter & description & value \\
\midrule[0.5pt]
$t_{max}$ & time limit (hours) & 0.5 or 2.0 \\
$k$ & elite set size    & 15 \\
$max\_no\_update$ & number of times without updating & 15 \\
$\theta$ & minimum support & 2 \\
$m$ & frequent pattern set size & 11 \\
$\lambda$ & tournament pool size & 3 \\
$\beta$ & length threshold & 0.75 \\
$max\_iter$ & number of iterations for BLS\tnote{$\star$} & 10000 \\
\bottomrule[0.75pt]
\end{tabular}
\begin{tablenotes}
     \item[$\star$] Other parameters of BLS adopt the default values provided in \cite{Benlic2013a}.
\end{tablenotes}
\end{threeparttable}
\end{table}

Given its stochastic nature, the proposed FPBS-QAP algorithm was independently ran 10 times on each test instance, which is a standard practice for solving QAP \cite{Benlic2013a,Benlic2015,Tosun2015,Acan2015}. Our assessment is based on the percentage deviation (PD) metrics that are widely used in the literature. The PD metrics measures the percentage deviation from the best-known value (BKV). For example, the best percentage deviation (BAD), the average percentage deviation (APD) and the worst percentage deviation (WPD), are respectively calculated according to:

\begin{equation}
	XPD = 100*\frac{X-BKV}{BKV}[\%]
\end{equation}

where $X \in \{B,A,W\}$ corresponds to the best objective value, average objective value and worst objective value achieved by an algorithm. The smaller the XPD value, the better the evaluated algorithm.

\subsubsection{Comparison of FPBS-QAP with BLS and BMA}
\label{SubSubSec:Benefit of the Proposed FPBS Algorithm}

To demonstrate the effectiveness of the FPBS-QAP algorithm, we first show a detailed comparison of FPBS-QAP with two main reference algorithms: BLS (breakout local search) \cite{Benlic2013a} and BMA (population-based memetic algorithm) \cite{Benlic2015}. This experiment was based on two motivations. First, BLS and BMA are among the best performing QAP algorithms in the literature. Second, the source codes of BLS and BMA are available to us, making it possible to make a fair comparision (using the same computing platform and stopping conditions). Third, both FPBS-QAP and BMA use BLS as their underlying optimization procedure, this comparison allows us to assess the added value of the data mining component of FPBS-QAP. For this experiment, we run FPBS-QAP and the two reference algorithms under two stopping conditions, i.e., a limit of $t_{max} = 30$ minutes (0.5 hour) and a limit of $t_{max} = 120$ minutes (2 hours). This allows us to study the behavior of the compared algorithms under short and long conditions.

The comparative performances of FPBS-QAP with BLS and BMA under $t_{max} = 30$ minutes and $t_{max} = 120$ minutes are presented in Table \ref{Tab:Comparisons on Between FPBS-QAP and BLS and BMA With T = 0.5h} and Table \ref{Tab:Comparisons on Between FPBS-QAP and BLS and BMA With T = 2h}, respectively. In these two tables, we report the BPD, APD, WPD values of each algorithm, and the average time in minutes ($T(m)$) to achieve the results. At the last row of each table, we also indicate the average value of each indicator. The smaller the value, the better the performance of an algorithm.

\begin{table*}[!ht]
\scriptsize\centering
\caption{Performance comparison of the proposed FPBS-QAP algorithm with BLS and BMA on 21 hard instances under $t_{max} = 30$ minutes. The number of times of reaching the best-known value over 10 runs is indicated in parentheses.}
\label{Tab:Comparisons on Between FPBS-QAP and BLS and BMA With T = 0.5h}
\begin{tabular}{ll|lccrclccrclccr}
\toprule[0.75pt]
\multicolumn{2}{c}{} & \multicolumn{4}{c}{BLS} && \multicolumn{4}{c}{BMA}&& \multicolumn{4}{c}{FPBS-QAP}\\
\cmidrule[0.5pt]{3-6} \cmidrule[0.5pt]{8-11} \cmidrule[0.5pt]{13-16}
Instance  & BKV	 & BPD & APD & WPD & $T(m)$ && BPD & APD & WPD & $T(m)$ && BPD & APD & WPD & $T(m)$\\
\midrule[0.5pt]
tai40a &3139370   &0.000(1) &0.067&0.074&5.5 &&0.000(1) &0.067&0.074&7.0 &&0.000(1) &0.067&0.074&6.4 \\
tai50a &4938796   &0.000(1) &0.181&0.364&14.5&&0.053(0) &0.216&0.317&12.8&&0.000(1) &0.279&0.415&17.3\\
tai60a &7205962   &0.273(0) &0.371&0.442&17.7&&0.165(0) &0.285&0.381&17.6&&0.165(0) &0.377&0.469&8.7 \\
tai80a &13499184  &0.474(0) &0.571&0.647&13.5&&0.468(0) &0.553&0.621&14.5&&0.430(0) &0.516&0.614&18.2\\
tai100a&21052466  &0.467(0) &0.566&0.630&16.4&&0.389(0) &0.510&0.623&19.0&&0.311(0) &0.402&0.553&13.4\\
tai50b &458821517 &0.000(10)&0.000&0.000&0.2 &&0.000(10)&0.000&0.000&0.2 &&0.000(10)&0.000&0.000&0.1 \\
tai60b &608215054 &0.000(10)&0.000&0.000&0.6 &&0.000(10)&0.000&0.000&0.3 &&0.000(10)&0.000&0.000&0.5 \\
tai80b &818415043 &0.000(10)&0.000&0.000&3.3 &&0.000(10)&0.000&0.000&1.8 &&0.000(10)&0.000&0.000&1.2 \\
tai100b&1185996137&0.000(9) &0.005&0.045&10.3&&0.000(6) &0.044&0.143&2.4 &&0.000(6) &0.040&0.100&4.0 \\
tai150b&498896643 &0.014(0) &0.219&0.390&14.5&&0.000(1) &0.137&0.316&21.5&&0.000(1) &0.191&0.321&20.5\\
sko72  &66256     &0.000(10)&0.000&0.000&4.4 &&0.000(9) &0.006&0.063&0.7 &&0.000(10)&0.000&0.000&1.4 \\
sko81  &90998     &0.000(9) &0.001&0.011&8.0 &&0.000(10)&0.000&0.000&3.5 &&0.000(10)&0.000&0.000&2.7 \\
sko90  &115534    &0.000(4) &0.023&0.095&9.5 &&0.000(9) &0.004&0.038&5.9 &&0.000(6) &0.015&0.038&4.2 \\
sko100a&152002    &0.000(4) &0.006&0.018&12.4&&0.000(8) &0.003&0.016&5.2 &&0.000(10)&0.000&0.000&7.7 \\
sko100b&153890    &0.000(8) &0.001&0.004&4.6 &&0.000(10)&0.000&0.000&8.5 &&0.000(10)&0.000&0.000&7.7 \\
sko100c&147862    &0.000(7) &0.001&0.004&10.2&&0.000(10)&0.000&0.000&6.9 &&0.000(10)&0.000&0.000&8.7 \\
sko100d&149576    &0.000(3) &0.004&0.009&13.5&&0.000(9) &0.007&0.066&6.6 &&0.000(10)&0.000&0.000&9.0 \\
sko100e&149150    &0.000(5) &0.002&0.005&13.7&&0.000(10)&0.000&0.000&6.5 &&0.000(7) &0.001&0.004&10.3\\
sko100f&149036    &0.000(5) &0.006&0.032&11.5&&0.000(6) &0.005&0.021&6.3 &&0.000(7) &0.003&0.021&4.4 \\
wil100 &273038    &0.000(6) &0.001&0.003&11.4&&0.000(10)&0.000&0.000&6.4 &&0.000(10)&0.000&0.000&9.5 \\
tho150 &8133398   &0.013(0) &0.091&0.128&11.9&&0.003(0) &0.021&0.067&19.9&&0.000(1) &0.051&0.123&24.2\\
\midrule[0.5pt]
avg.	   &          &0.059    &0.101&0.138&9.8 &&0.051    &0.095&0.139&\textbf{8.2} &&\textbf{0.043}    &\textbf{0.092}&\textbf{0.130}&8.6 \\
\bottomrule[0.75pt]
\end{tabular}
\end{table*}

From Table \ref{Tab:Comparisons on Between FPBS-QAP and BLS and BMA With T = 0.5h}, we observe that FPBS-QAP achieves the best performance compared to the algorithms BLS and BMA under $t_{max} = 30$ minutes. First, FPBS-QAP achieves all best-known values except two cases while BLS and BMA fail to find the best-known values for five instances. Second, FPBS-QAP is able to reach the best-known values of the two largest instances, i.e., tai150b and tho150 within the given computing time. BLS fails to find the best-known values of tai150b and tho150 within the time limit of 30 minutes (it can find these values only under a very long time limit of $t_{max} = 10$ hours). BMA performs better than BLS by attaining the best-known value of tai150b, but still fails on tho150. The BPD value of FPBS-QAP for the 21 benchmark instances is only 0.043$\%$, which is smaller than 0.059$\%$ of BLS, and 0.051$\%$ of BMA respectively. Similar observations can also be found for the APD and WPD indicators. It is worth noting that FPBS-QAP needs less time to achieve these (better) results than BLS, while it consumes nearly the same computing time as BMA.

When a long time limit of $t_{max} = 120$ minutes is allowed, our FPBS-QAP algorithm is able to achieve even better results. As we see from Table \ref{Tab:Comparisons on Between FPBS-QAP and BLS and BMA With T = 2h}, the best-known values are obtained with a higher success rate compared to the results under $t_{max} = 30$ minutes in Table \ref{Tab:Comparisons on Between FPBS-QAP and BLS and BMA With T = 0.5h}. The average BPD value of FPBS-QAP is $0.028\%$, which is the smallest compared to $0.037\%$ of BMA, and $0.038\%$ of BLS. FPBS-QAP also achieves the smallest average APD value and average WPD value. As to the computing times, FPBS-QAP requires on average 22.0 minutes to reach its best solution, which is the shortest time among the compared algorithms (32.2 minutes for BLS, and 23.1 minutes for BMA).

\begin{table*}[!ht]
\scriptsize\centering
\caption{Performance comparison of the proposed FPBS-QAP algorithm with BLS and BMA on 21 hard instances under $t_{max} = 120$ minutes. The number of times of reaching the best-known value over 10 runs is indicated in parentheses.}
\label{Tab:Comparisons on Between FPBS-QAP and BLS and BMA With T = 2h}
\begin{tabular}{ll|lccrclccrclccr}
\toprule[0.75pt]
\multicolumn{2}{c}{} & \multicolumn{4}{c}{BLS} && \multicolumn{4}{c}{BMA}&& \multicolumn{4}{c}{FPBS-QAP}\\
\cmidrule[0.5pt]{3-6} \cmidrule[0.5pt]{8-11} \cmidrule[0.5pt]{13-16}
Instance  & BKV	 & BPD & APD & WPD & $T(m)$ && BPD & APD & WPD & $T(m)$ && BPD & APD & WPD & $T(m)$\\
\midrule[0.5pt]
tai40a &3139370   &0.000(7) &0.022&0.074&40.7&&0.000(5) &0.037&0.074&28.9&&0.000(7) &0.022&0.074&52.5\\
tai50a &4938796   &0.000(1) &0.100&0.245&47.0&&0.000(3) &0.098&0.291&38.8&&0.000(2) &0.107&0.231&67.8\\
tai60a &7205962   &0.036(0) &0.233&0.331&73.9&&0.165(0) &0.221&0.352&34.7&&0.000(1) &0.216&0.300&60.0\\
tai80a &13499184  &0.416(0) &0.502&0.587&58.0&&0.332(0) &0.428&0.505&69.9&&0.313(0) &0.451&0.618&55.2\\
tai100a&21052466  &0.335(0) &0.460&0.560&58.4&&0.223(0) &0.370&0.511&59.9&&0.280(0) &0.378&0.466&36.1\\
tai50b &458821517 &0.000(10)&0.000&0.000&0.2 &&0.000(10)&0.000&0.000&0.1 &&0.000(10)&0.000&0.000&0.2 \\
tai60b &608215054 &0.000(10)&0.000&0.000&0.3 &&0.000(10)&0.000&0.000&0.4 &&0.000(10)&0.000&0.000&0.4 \\
tai80b &818415043 &0.000(10)&0.000&0.000&3.9 &&0.000(10)&0.000&0.000&2.0 &&0.000(10)&0.000&0.000&1.4 \\
tai100b&1185996137&0.000(10)&0.000&0.000&6.1 &&0.000(6) &0.040&0.100&8.1 &&0.000(10)&0.000&0.000&2.9 \\
tai150b&498896643 &0.001(0) &0.040&0.183&42.7&&0.055(0) &0.213&0.414&56.0&&0.000(5) &0.099&0.313&46.4\\
sko72  &66256     &0.000(10)&0.000&0.000&3.0 &&0.000(10)&0.000&0.000&0.7 &&0.000(10)&0.000&0.000&4.8 \\
sko81  &90998     &0.000(10)&0.000&0.000&10.3&&0.000(10)&0.000&0.000&3.2 &&0.000(10)&0.000&0.000&3.3 \\
sko90  &115534    &0.000(10)&0.000&0.000&19.6&&0.000(9) &0.004&0.038&3.6 &&0.000(9) &0.004&0.038&2.4 \\
sko100a&152002    &0.000(10)&0.000&0.000&51.0&&0.000(8) &0.003&0.016&28.6&&0.000(10)&0.000&0.000&8.5 \\
sko100b&153890    &0.000(10)&0.000&0.000&21.6&&0.000(10)&0.000&0.000&11.0&&0.000(10)&0.000&0.000&5.8 \\
sko100c&147862    &0.000(10)&0.000&0.000&22.4&&0.000(10)&0.000&0.000&7.1 &&0.000(10)&0.000&0.000&8.7 \\
sko100d&149576    &0.000(6) &0.001&0.005&38.5&&0.000(10)&0.000&0.000&12.6&&0.000(10)&0.000&0.000&16.2\\
sko100e&149150    &0.000(10)&0.000&0.000&44.2&&0.000(10)&0.000&0.000&5.3 &&0.000(10)&0.000&0.000&12.2\\
sko100f&149036    &0.000(7) &0.002&0.005&40.2&&0.000(8) &0.001&0.005&23.7&&0.000(6) &0.002&0.005&4.0 \\
wil100 &273038    &0.000(9) &0.000&0.002&28.9&&0.000(10)&0.000&0.000&6.9 &&0.000(10)&0.000&0.000&16.4\\
tho150 &8133398   &0.009(0) &0.056&0.116&64.2&&0.000(1) &0.036&0.065&83.9&&0.000(3) &0.008&0.064&57.4\\
\midrule[0.5pt]
avg.	   &  &0.038 & 0.067 & \textbf{0.100} & 32.2 && 0.037 & 0.069 & 0.113 &23.1 && \textbf{0.028} & \textbf{0.061} & \textbf{0.100} & \textbf{22.0}\\
\bottomrule[0.75pt]
\end{tabular}
\end{table*}

In summary, our FPBS-QAP algorithm competes favorably with the two best-performing QAP algorithms (i.e., BLS and BMA) in terms of both solution quality and computing time. The computational results demonstrate the effectiveness of the proposed FPBS-QAP algorithm, and further shows the usefulness of using frequent patterns mined from high-quality solutions to guide the search for an effective exploration of the solution space.

\subsubsection{Comparison with four other state-of-the-art algorithms}
\label{SubSubSec:Comparison With the State-of-the-art Algorithms}

We now extend our experimental study by comparing FPBS-QAP with four other very recent state-of-the-art QAP algorithms in the literature. 

\begin{itemize}
	\item Parallel hybrid algorithm (PHA) \cite{Tosun2015} (2015) used the MPI libraries and was implemented on a high-performance cluster with 46 nodes, a total RAM of 736 GB and a total disk capacity of 6.5 TB configured in a high-performance RAID. Each node includes 2 CPUs (4 cores per CPU) and 16 GB of RAM.
	\item Two-stage memory powered great deluge algorithm (TMSGD) \cite{Acan2015} (2015) was implemented on a personal computer with 2.1 GHz and 8 GB RAM. The algorithm was stopped when the number of fitness evaluations reaches $20000*n$ ($n$ is the instance size).
	\item Parallel multi-start hyper-heuristic algorithm (MSH) \cite{Dokeroglu2016} (2016) was implemented on the same high performance cluster as the above PHA algorithm.
	\item Breakout local search using OpenMP (BLS-OpenMP) \cite{Aksan2017} (2017) was implemented on OpenMP (i.e., an API for shared-memory parallel computations that runs on multi-core computers.) and was executed on a personal computer with an Intel Core i7-6700 CPU 3.4 GHZ with 4 cores and 16 GB RAM. It is possible to execute 8 logical processors on this computer.
\end{itemize}

One notices that three of these four recent QAP algorithms are implemented and run on parallel machines. Moreover, their results have been obtained on different computing platforms, with different stopping conditions. The comparison shown in this section is provided mainly for indicative purposes. Still, the comparison provides interesting indications on the performance of the proposed algorithm relative to these state-of-the-art algorithms. Moreover, the availability of our FPBS-QAP code makes it possible for researchers to make fair comparisons with FPBS-QAP (see footnote 3).

Table \ref{Tab:Comparisons With The State-of-the-art Algorithms} presents the comparative results between the proposed FPBS-QAP algorithm and the four reference algorithms. Like \cite{Acan2015,Dokeroglu2016,Aksan2017}, we adopt the APD indicator (defined in Section \ref{SubSubSec:Experimental Settings}) for this comparative study and indicate the running time ($T(m)$) as an additional indicator, which should be interpreted with cautions for the reasons raised above. For completeness, we also include the results of BLS and BMA from Table \ref{Tab:Comparisons With The State-of-the-art Algorithms}. In the last row of the table, we again indicate the average value of each indicator. 

\begin{table*}[!ht]
\scriptsize\centering
\caption{Comparative performance between the FPBS-QAP algorithm and state-of-the-art algorithms on hard instances in terms of the APD value. Computational time are given in minutes for indicative purposes.}
\label{Tab:Comparisons With The State-of-the-art Algorithms}
\begin{threeparttable}
\begin{tabular}{ll|crcrcrcrcrcrcr}
\toprule[0.75pt]
\multicolumn{2}{c}{} & \multicolumn{2}{c}{BLS\tnote{$\star$}} & \multicolumn{2}{c}{PHA\tnote{$\circ$}}& \multicolumn{2}{c}{BMA\tnote{$\star$}} & \multicolumn{2}{c}{TMSGD} & \multicolumn{2}{c}{MSH\tnote{$\circ$}} & \multicolumn{2}{c}{BLS-OpenMP\tnote{$\circ$}} & \multicolumn{2}{c}{FPBS-QAP}\\
\cmidrule[0.5pt]{3-4} \cmidrule[0.5pt]{5-6} \cmidrule[0.5pt]{7-8} \cmidrule[0.5pt]{9-10} \cmidrule[0.5pt]{11-12} \cmidrule[0.5pt]{13-14} \cmidrule[0.5pt]{15-16} 
Instance  & BKV	 & APD & $T(m)$ & APD & $T(m)$ & APD & $T(m)$ & APD & $T(m)$ & APD & $T(m)$ & APD & $T(m)$ & APD & $T(m)$\\
\midrule[0.5pt]
tai40a &3139370&0.022&40.7&0.000&10.6&0.037&28.9&0.261&27.8&0.261&30.0&0.000&32.2&0.022&52.5\\
tai50a &4938796&0.100&47.0&0.000&12.7&0.098&38.8&0.276&41.1&0.165&37.5&0.000&68.2&0.107&67.8\\
tai60a &7205962&0.233&73.9&0.000&19.6&0.221&34.7&0.448&78.9&0.270&45.0&0.000&107.9&0.216&60.0\\
tai80a &13499184&0.502&58.0&0.644&40.0&0.428&69.9&0.832&111.3&0.530&60.0&0.504&236.0&0.451&55.2\\
tai100a&21052466&0.460&58.4&0.537&71.9&0.370&59.9&0.874&138.3&0.338&75.0&0.617&448.5&0.378&36.1\\
tai50b &458821517&0.000&0.2&0.000&5.8&0.000&0.1&0.005&10.2&0.000&3.0&0.000&0.7&0.000&0.2\\
tai60b &608215054&0.000&0.3&0.000&9.5&0.000&0.4&0.000&33.6&0.000&3.2&0.000&18.6&0.000&0.4\\
tai80b &818415043&0.000&3.9&0.000&27.7&0.000&2.0&0.025&0.0&0.000&4.0&0.000&218.1&0.000&1.4\\
tai100b&1185996137&0.000&6.1&0.000&42.5&0.040&8.1&0.028&72.6&0.000&5.0&0.000&160.8&0.000&2.9\\
tai150b&498896643&0.040&42.7&0.026&177.4&0.213&56.0&0.051&258.0&*&*&*&*&0.099&46.4\\
sko72  &66256&0.000&3.0&0.000&33.6&0.000&0.7&0.007&38.0&0.000&3.6&0.000&1.8&0.000&4.8\\
sko81  &90998&0.000&10.3&0.000&39.9&0.000&3.2&0.019&57.1&0.000&4.1&0.000&2.4&0.000&3.3\\
sko90  &115534&0.000&19.6&0.000&40.5&0.004&3.6&0.031&93.8&0.000&4.5&0.000&3.3&0.004&2.4\\
sko100a&152002&0.000&51.0&0.000&41.7&0.003&28.6&0.029&153.2&0.003&75.0&0.000&29.8&0.000&8.5\\
sko100b&153890&0.000&21.6&0.000&42.3&0.000&11.0&0.015&164.3&0.004&75.0&0.000&8.5&0.000&5.8\\
sko100c&147862&0.000&22.4&0.000&42.2&0.000&7.1&0.013&154.5&0.003&75.0&0.000&4.3&0.000&8.7\\
sko100d&149576&0.001&38.5&0.000&41.9&0.000&12.6&0.017&148.9&0.004&75.0&0.000&12.9&0.000&16.2\\
sko100e&149150&0.000&44.2&0.000&42.5&0.000&5.3&0.016&146.1&0.000&75.0&0.000&4.3&0.000&12.2\\
sko100f&149036&0.002&40.2&0.000&42.0&0.001&23.7&0.013&153.4&0.000&75.0&0.000&17.1&0.002&4.0\\
wil100 &273038&0.000&28.9&0.000&42.0&0.000&6.9&0.008&155.1&*&*&*&*&0.000&16.4\\
tho150 &8133398&0.056&64.2&0.009&177.4&0.036&83.9&0.039&512.8&*&*&*&*&0.008&57.4\\
\midrule[0.5pt]
avg. & &0.067 & 32.2 & \textbf{0.058} & 47.8 & 0.069 & 23.1 & 0.143 & 121.4 & 0.088 & 40.3 & 0.062 & 76.4 & 0.061 & \textbf{22.0}\\
\bottomrule[0.75pt]
\end{tabular}
\begin{tablenotes}
     \item[$\star$] The results of BLS and BMA were obtained by running the programs on our computer with $t_{max} = 120$ minutes. These results are slightly different from the results reported in \cite{Benlic2013a,Benlic2015}.
     \item[$\circ$] PHA, MSH and BLS-OpenMP are parallel algorithms that were run on high-performance platforms under various stopping conditions.
\end{tablenotes}
\end{threeparttable}
\end{table*}

From Table \ref{Tab:Comparisons With The State-of-the-art Algorithms}, we observe that our FPBS-QAP algorithm achieves a highly competitive performance compared to these state-of-the-art algorithms. The average APD value of FPBS-QAP is $0.061\%$, which is only slightly worse than $0.058\%$ of the parallel PHA algorithm and better than all remaining reference algorithms. Moreover, even if PHA was run on a parallel high-performance computing platform, its average computing time ($\geq 177.4$ minutes) is almost three times of the time required by FPBS-QAP ($\leq 67.8$ minutes) to obtain very similar results. Note that the results of three instances, including the two hardest and largest instances tai150b and tho150, are not reported for BLS-OpenMP. The APD value of BLS-OpenMP is computed for the remaining 18 instances. Importantly, our algorithm requires the least time to achieve the best results, and its average time is only 22.0 minutes. These observations show that our FPBS-QAP algorithm is highly competitive compared to the state-of-the-art algorithms in terms of solution quality and computing time.

\section{Analysis and discussion}
\label{Sec:Analysis and Discussion}

In this section, we perform additional experiments to gain understandings of the proposed FPBS algorithm including the rationale behind the solution construction based on frequent patterns, the effectiveness of the solution construction based on mined frequent patterns, and the impact of the number of the largest patterns $m$ on the performance of the proposed algorithm.

\subsection{Rationale behind the solution construction based on mined patterns}
\label{SubSec:Rationale Behind the Solution Construction Based on Mined Patterns}

To explain the rationale behind the solution construction based on mined frequent patterns, we analyze the structural similarity between high-quality solutions in the elite set, and the length distribution of the frequent patterns mined from the elite set. Given two high-quality solutions $\pi^s$ and $\pi^t$, we define their similarity as follows.

\begin{equation}\label{Equ:Solution Similarity}
	sim(\pi^s, \pi^t) = \frac{|\pi^s \cap \pi^t|}{n}
\end{equation}

where $\pi^s \cap \pi^t$ is the set of common elements shared by $\pi^s$ and $\pi^t$. The larger the similarity between two solutions, the more common elements they share.

\begin{figure}[!htbp]
\centering
\includegraphics[width=3.5in]{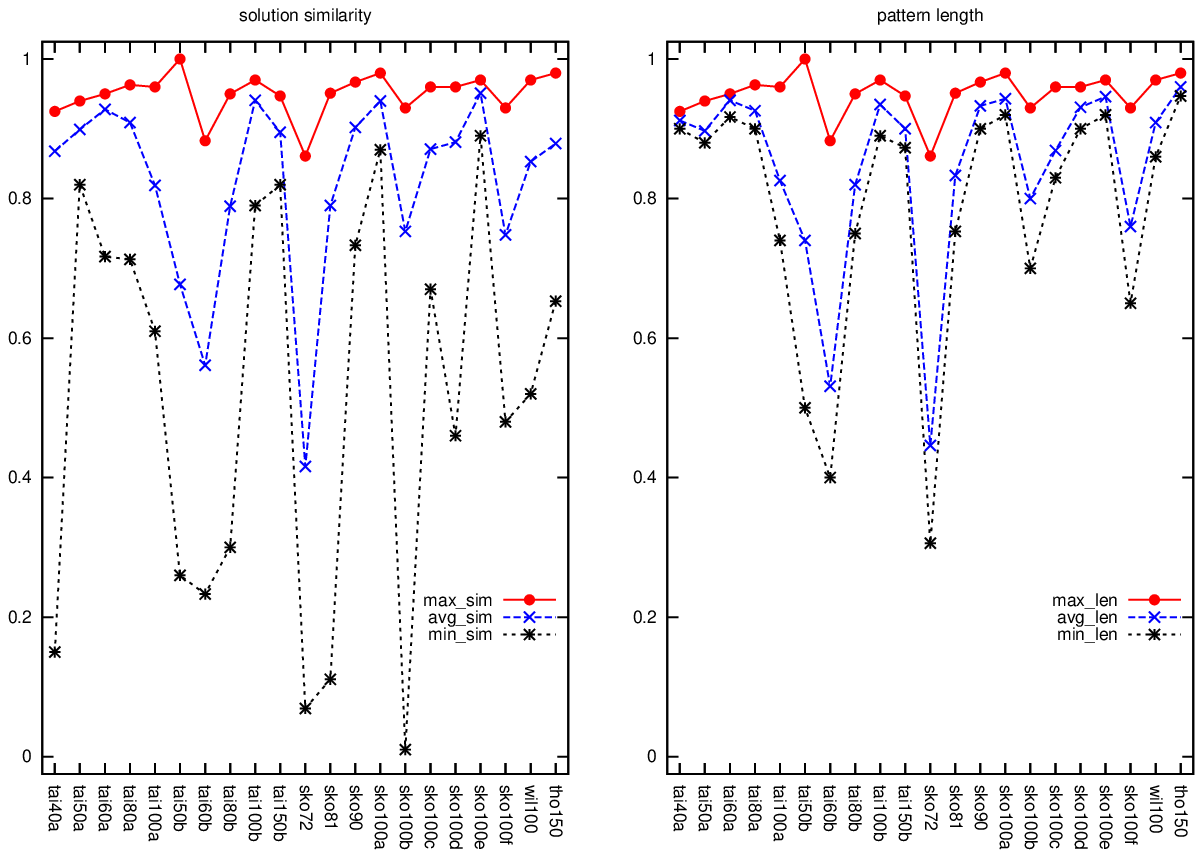}
\caption{Solution similarity between high-quality solutions (left sub-figure) and length distribution of the mined patterns (right sub-figure).}
\label{Fig:Structural Similarity Among the Elite Solutions}
\end{figure}

As we mentioned above, a mined frequent pattern represents a set of identical elements shared by two or more solutions under a given minimum support $\theta$. A frequent pattern can be directly converted to a partial solution, thus we define the length of a pattern $p$ as follows.

\begin{equation} \label{Equ:Pattern Length}
	len(p) = \frac{|p|}{n}
\end{equation}

where the length of a pattern is the proportion of the number of identical elements over the total number of elements. A larger pattern length indicates thus more shared elements. The solution similarity defined in Def. (\ref{Equ:Solution Similarity}) can be considered as a special case of the pattern length defined in Def. (\ref{Equ:Pattern Length}). Specifically, when the support value of a mined pattern equals 2, the pattern is simplified as the set of common elements shared by only two solutions. This is further confirmed according to the results reported in Figure \ref{Fig:Structural Similarity Among the Elite Solutions}, where the curve of the maximum solution similarity (left sub-figure) is exactly the same as the curve of the maximum length (right sub-figure).

In this experiment, we solved each benchmark instance with a time limit of $t_{max} = 30$ minutes. To analyze the solution similarity of high-quality solutions stored in the elite set according to Eq. (\ref{Equ:Solution Similarity}), we calculate the length distribution of a set of frequent patterns mined from the elite set according to Eq. (\ref{Equ:Pattern Length}). The results of the similarity between high-quality solutions and the length distribution of the mined frequent patterns are presented in Figure \ref{Fig:Structural Similarity Among the Elite Solutions}. 

In Figure \ref{Fig:Structural Similarity Among the Elite Solutions}, we report the maximum value, average value, and minimum value of the solution similarity and the pattern length, respectively. We can clearly observe that there is a high similarity between the high-quality solutions. Specifically, for all instances, the maximum solution similarity is larger than 0.9. Also, the average solution similarities between any two high-quality solutions are larger than 0.5 except for sko72, for which the average solution similarity is about 0.4. A more significant observation can be derived based on the lengths of the mined patterns showed in the right sub-figure. The high structural similarities between the high-quality solutions provide the rationale behind our solution construction based on mined patterns.

\subsection{Effectiveness of the solution construction based on frequent pattern}
\label{SubSec:Effectiveness of the Solution Construction Based on Frequent Pattern}

The frequent pattern based solution construction method is a good alternative to the general crossover operator in evolutionary algorithms and memetic algorithms. To demonstrate the effectiveness of the solution construction using frequent patterns, we compare this approach with the general crossover operator within the framework of our FPBS-QAP algorithm. In this experiment, we compared FPBS-QAP with its alternative version FPBS-QAP$_0$ where the frequent pattern based solution construction of FPBS-QAP was replaced by the standard uniform crossover operator used in \cite{Benlic2015}. We ran both algorithms on each benchmark instance 10 times with a time limit of $t_{max} = 30$ minutes. The comparative results between FPBS-QAP and FPBS-QAP$_0$ are summarized in Table \ref{Tab:Comparisons on Between FPBS-QAP_0 and FPBS-QAP With T = 0.5h}.

\begin{table*}[!ht] 
\scriptsize\centering
\caption{Comparisons between FPBS-QAP$_0$ and FPBS-QAP on hard instances under the time limit of $t_{max} = 30$ minutes. The success rate of reaching the best-known value over 10 runs is indicated in parentheses.}
\label{Tab:Comparisons on Between FPBS-QAP_0 and FPBS-QAP With T = 0.5h}
\begin{threeparttable}
\begin{tabular}{ll|lccrclccr}
\toprule[0.75pt]
\multicolumn{2}{c}{} & \multicolumn{4}{c}{FPBS-QAP$_0$\tnote{$\star$}} && \multicolumn{4}{c}{FPBS-QAP}\\
\cmidrule[0.5pt]{3-6} \cmidrule[0.5pt]{8-11}
Instance  & BKV	 & BPD & APD & WPD & $T(m)$ && BPD & APD & WPD & $T(m)$\\
\midrule[0.5pt]
tai40a&3139370&0.000(2)&0.059&0.074&7.4&&0.000(1)&0.067&0.074&6.4\\
tai50a&4938796&0.241(0)&0.318&0.392&11.8&&0.000(1)&0.279&0.415&17.3\\
tai60a&7205962&0.164(0)&0.334&0.486&13.1&&0.165(0)&0.377&0.469&8.7\\
tai80a&13499184&0.446(0)&0.533&0.622&15.7&&0.430(0)&0.516&0.614&18.2\\
tai100a&21052466&0.316(0)&0.466&0.615&16.8&&0.311(0)&0.402&0.553&13.4\\
tai50b&458821517&0.000(10)&0.000&0.000&0.1&&0.000(10)&0.000&0.000&0.1\\
tai60b&608215054&0.000(10)&0.000&0.000&0.3&&0.000(10)&0.000&0.000&0.5\\
tai80b&818415043&0.000(10)&0.000&0.000&1.5&&0.000(10)&0.000&0.000&1.2\\
tai100b&1185996137&0.000(8)&0.018&0.100&3.1&&0.000(6)&0.040&0.100&4.0\\
tai150b&498896643&0.000(1)&0.204&0.358&20.2&&0.000(1)&0.191&0.321&20.5\\
sko72&66256&0.000(9)&0.006&0.063&2.2&&0.000(10)&0.000&0.000&1.4\\
sko81&90998&0.000(10)&0.000&0.000&5.1&&0.000(10)&0.000&0.000&2.7\\
sko90&115534&0.000(9)&0.004&0.038&4.1&&0.000(6)&0.015&0.038&4.2\\
sko100a&152002&0.000(9)&0.002&0.016&7.0&&0.000(10)&0.000&0.000&7.7\\
sko100b&153890&0.000(8)&0.001&0.004&6.0&&0.000(10)&0.000&0.000&7.7\\
sko100c&147862&0.000(10)&0.000&0.000&6.0&&0.000(10)&0.000&0.000&8.7\\
sko100d&149576&0.000(10)&0.000&0.000&9.5&&0.000(10)&0.000&0.000&9.0\\
sko100e&149150&0.000(10)&0.000&0.000&7.6&&0.000(7)&0.001&0.004&10.3\\
sko100f&149036&0.000(7)&0.002&0.005&3.6&&0.000(7)&0.003&0.021&4.4\\
wil100&273038&0.000(9)&0.000&0.002&6.0&&0.000(10)&0.000&0.000&9.5\\
tho150&8133398&0.002(0)&0.022&0.080&19.1&&0.000(1)&0.051&0.123&24.2\\
\midrule[0.5pt]
avg. & & 0.056 & 0.094 & 0.136 & \textbf{7.9} && \textbf{0.043} & \textbf{0.092} & \textbf{0.130} & 8.6\\
\bottomrule[0.75pt]
\end{tabular}
\begin{tablenotes}
     \item[$\star$] FPBS-QAP$_0$ can also be considered as an alternative version of BMA \cite{Benlic2015} by removing the mutation procedure.
\end{tablenotes}
\end{threeparttable}
\end{table*}

Table \ref{Tab:Comparisons on Between FPBS-QAP_0 and FPBS-QAP With T = 0.5h} indicates that FPBS-QAP performs better than FPBS-QAP$_0$. FPBS-QAP is able to achieve a better or the same BPD value on all instances except tai60a. For tai60a, the BPD value of FPBS-QAP is $0.165\%$, which is only marginally worse than $0.164\%$ achieved by FPBS-QAP$_0$. The average BPD value of FPBS-QAP is also better than that of FPBS-QAP$_0$, i.e., $0.043\% < 0.056\%$. Finally, FPBS-QAP achieves better results in terms of the average APD value and the average WPD value. To achieve these results, the average run time of FPBS-QAP$_0$ is slightly shorter than that of FPBS-QAP (i.e., $7.9 < 8.6$ minutes). This can be explained by the fact that FPBS-QAP needs to execute a frequent pattern mining procedure during the search. These observations confirms the interest of the solution construction method using mined frequent patterns.

\subsection{Impact of the number of the largest frequent patterns $m$}
\label{SubSec:Impact of the Number of the Largest Patterns}

\begin{figure}[!htbp]
\centering
\includegraphics[width=3.8in]{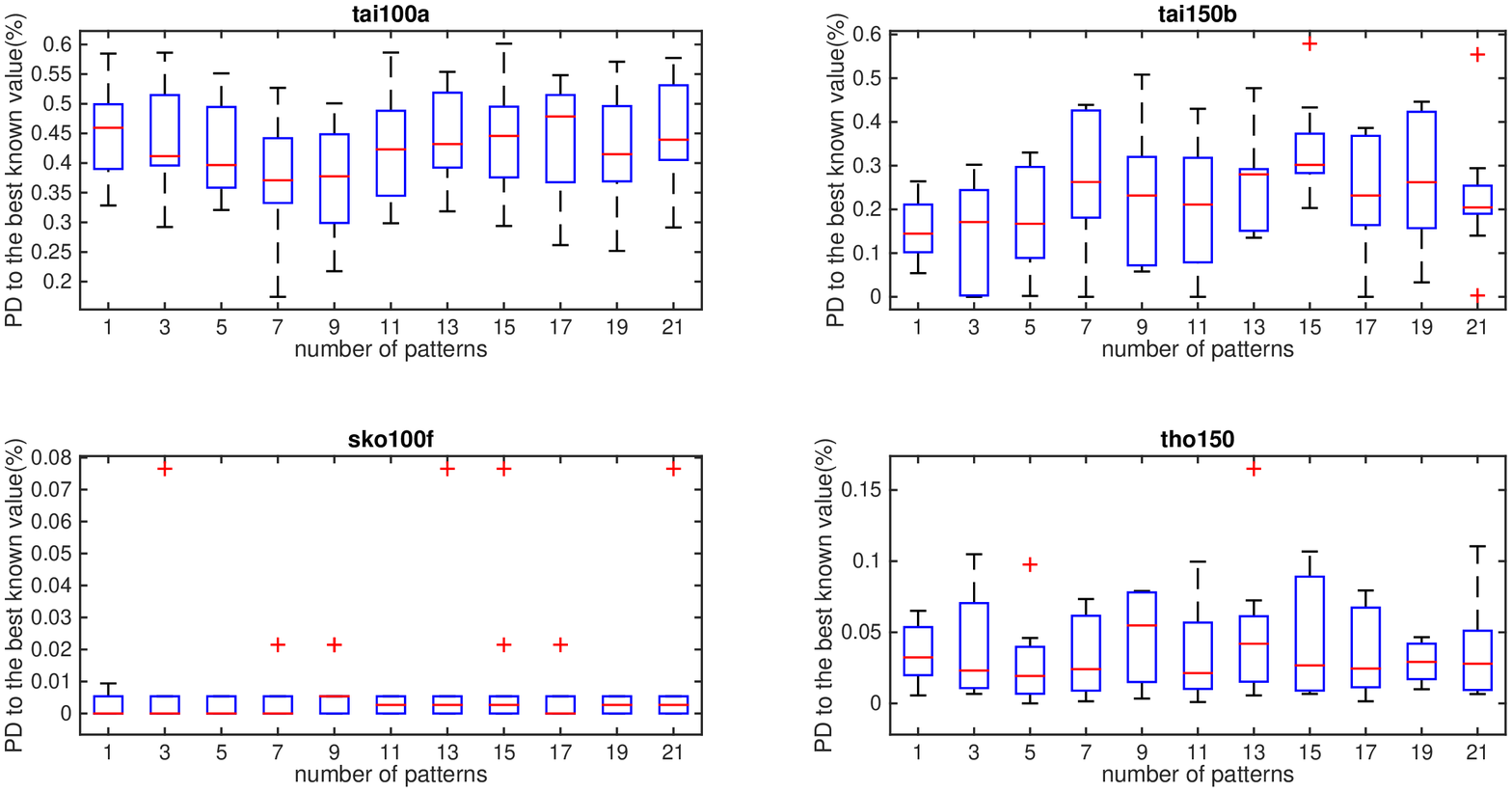}
\caption{Impact of the number of the largest frequent patterns $m$. Box and whisker plots corresponding to 10 different values of $m \in \{1,3,\ldots,21\}$ in terms of the percentage deviation (PD) to the best-known value.}
\label{Fig:Boxplots With Different NbrPatterns}
\end{figure}

The number of the longest frequent patterns $m~(m \geqslant 1)$ influences the diversity of the new solutions constructed by the solution construction method using mined frequent patterns. To investigate the impact of this parameter, we varied the values of $m$ within a reasonable range and compared their performances. The box and whisker plots showed in Figure \ref{Fig:Boxplots With Different NbrPatterns} are obtained by considering ten different values $m \in \{1,3,\ldots,21\}$. The experiments were conducted on four representative instances selected from different families (tai100a, tai150b, sko100f and tho150). For each $m$ value and each instance, we ran the algorithm 10 times with the stopping condition of $t_{max} = 30$ minutes.

In Figure \ref{Fig:Boxplots With Different NbrPatterns}, X-axis indicates the different values for the number of the largest frequent patterns $m$ and Y-axis shows the performance (i.e., the percentage deviation to the best-known value). We observe that the performance of the FPBS-QAP algorithm strongly depends on the $m$ value except for sko100f. FPBS-QAP achieves a good performance when the number of the largest pattern $m$ is fixed to 11. This justifies the default value for $m$ shown in Table \ref{Tab:Parameter Settings}. 

To tune parameters $\beta$ and $max\_no\_update$, we used the same method and chosen $\beta = 0.75$, $\max\_no\_update = 15$ as their default values. It is possible that FPBS-QAP improves its performance when its parameters are tuned for each specific problem instance. 

\section{Conclusions and further work}
\label{Sec:Conclusions and Further Work}

In this paper, we proposed a general-purpose optimization approach called frequent pattern based search (FPBS). The proposed approach relies on a data mining procedure to mine frequent patterns from high-quality solutions collected during the search. The mined patterns are then used to create new starting solutions for further improvements. By iterating the pattern mining phase and the optimization phase, FPBS is designed to ensure an effective exploration of the combinatorial search space. 

The viability of the proposed approach was verified on the well-known quadratic assignment problem. Extensive computational results on  popular QAPLIB benchmarks showed that FPBS performs remarkably well compared to very recent and state-of-the-art algorithms both in terms of solution quality and computational efficiency. Specifically, our approach is able to find the best-known objective values for all the benchmark instances except tai80a and tai100a within a time limit of 0.5 hour or 2 hours. To the best of our knowledge, very few QAP algorithms can achieve such a performance. Furthermore, we performed additional experiments to investigate three key issues of the proposed FPBS algorithm.

As future work, three directions can be followed. First, this study focused on exploring maximal frequent itemsets. However, there are other interesting patterns available in the area of pattern mining. It is worth studying alternative patterns like sequential patterns and graph patterns. Second, FPBS is a general-purpose approach, it would be interesting to investigate its interest to solve other optimization problems, particularly other permutation problems (e.g., linear ordering problem and traveling salesman problem) and subset selection problems (e.g., diversity or dispersion problems and critical node problems).

\ifCLASSOPTIONcaptionsoff
  \newpage
\fi

\end{document}